\documentclass[letterpaper,10pt,conference]{ieeeconf}%

\usepackage{hyperref} 
\usepackage{graphicx}
\usepackage{subfigure}
\usepackage{float}
\usepackage{epstopdf}
\usepackage{amsthm}
\usepackage{algorithm}
\usepackage[noend]{algpseudocode}
\usepackage[cmex10]{amsmath}
\usepackage{amsfonts,amssymb}
\usepackage{times}
\usepackage{bm}
\usepackage{bbm}
\usepackage{color}
\usepackage{graphicx}
\usepackage{amsfonts}
\usepackage{cite}
\usepackage{mathtools}
\usepackage{amssymb}%
\usepackage{tabu}
\usepackage{color}
\providecommand{\U}[1]{\protect\rule{.1in}{.1in}}
\IEEEoverridecommandlockouts
\theoremstyle{definition}

\renewcommand{\baselinestretch}{0.98}
\usepackage{tcolorbox}
\usepackage{lipsum}
\usepackage{colortbl}
\tcbuselibrary{skins,breakable}
\usetikzlibrary{shadings,shadows}
\usepackage{xcolor}
    {\endtcolorbox}
\makeatletter
\def\BState{\State\hskip-\ALG@thistlm}
\makeatother

\newcommand{\real}{\mathbb{R}}
\newcommand*{\myDots}{\ifmmode\mathellipsis\else.\kern-0.13em.\kern-0.13em.\fi}

\title{{\LARGE \textbf{Simultaneous Localization and Parameter Estimation for\\ Single Particle Tracking via Sigma Points based EM}}\vspace{-1mm}}

\author{Ye Lin$^{1}$ and Sean B. Andersson$^{1,2}$ \\
$^1$Division of Systems Engineering, $^2$Department of Mechanical Engineering \\
Boston University, Boston, MA 02215, USA \\
\{yelin,sanderss\}@bu.edu}

\begin{document}
\maketitle

\begin{abstract}
Single Particle Tracking (SPT) is a powerful class of tools for analyzing the dynamics of individual biological macromolecules moving inside living cells. The acquired data is typically in the form of a sequence of camera images that are then post-processed to reveal details about the motion. In this work, we develop an algorithm for jointly estimating both particle trajectory and motion model parameters from the data. Our approach uses Expectation Maximization (EM) combined with an Unscented Kalman filter (UKF) and an Unscented Rauch-Tung-Striebel smoother (URTSS), allowing us to use an accurate, nonlinear model of the observations acquired by the camera. Due to the shot noise characteristics of the photon generation process, this model uses a Poisson distribution to capture the measurement noise inherent in imaging. In order to apply a UKF, we first must transform the measurements into a model with additive Gaussian noise. We consider two approaches, one based on variance stabilizing transformations (where we compare  the Anscombe and Freeman-Tukey transforms) and one on a Gaussian approximation to the Poisson distribution. Through simulations, we demonstrate efficacy of the approach and explore the differences among these measurement transformations.

\end{abstract}


\section{INTRODUCTION}

\label{sec:intro}
Single particle tracking (SPT) is an important class of techniques for studying the motion of single biological macromolecules. With its ability to localize particles with an accuracy far below the diffraction limit of light and the ability to track the trajectory across time, SPT continues to be an invaluable tool in understanding biology at the nanometer-scale. Under the standard approach, the images are post-processed individually to determine the location of each particle in the frame and then these positions are linked across frames to create a trajectory \cite{chenouard2014objective}. This trajectory is then further analyzed, typically by fitting the Mean Square Displacement (MSD) curve to an appropriate motion model to determine parameters such as diffusion coefficients. Regardless of the algorithms used, the paradigm separates trajectory estimation from model parameter identification, though it is clear that these two problems are coupled. In addition, the techniques for model parameter estimation assume a simple linear observation of the true particle position corrupted by additive white Gaussian noise. The actual data, however, are intensity measurements from a CCD camera. These measurements are well modeled as Poisson-distributed random variables with a rate that depends on the true location of the particle as well as on experimental realities, including background noise and details of the optics used in the instrument. This already nonlinear model becomes even more complicated at the low signal intensities common to SPT data where noise models specific to the type of camera being used become important \cite{krull2014divide, lin2017algorithmic}.

To handle such nonlinearities, one of the authors previously introduced an approach based on nonlinear system identification that uses Expectation Maximization (EM) combined with particle filtering and smoothing \cite{ashley2015method}. This general approach can handle nearly arbitrary nonlinearities in both the motion and observation models and has been shown to work as well as current state-of-the-art methods in the simple setting of 2-D diffusion. However, a major drawback of this approach is the computational complexity of the particle filtering scheme. In this paper we address this issue by replacing the particle-based methods with an Unscented Kalman filter (UKF) and Unscented Rauch-Tung-Striebel smoother (URTSS) \cite{van2004sigma,sarkka2013bayesian}. This Sigma Points based EM scheme, which we simply term as Unscented EM (U-EM), is significantly cheaper to implement, allowing it to be applied to larger data sets and for more complicated models. This reduction in complexity comes, of course, at the cost of generality in the posterior distribution describing the position of the particle at each time point since the UKF-URTSS approximates this distribution as a Gaussian while the particle-based approaches can represent other distributions \cite{sarkka2013bayesian}. 

One of the challenges in applying the UKF is that it assumes Gaussian noise in both the state update and measurement equations. In this work we focus on diffusion to focus the discussion on a concrete setting. As the corresponding dynamic model is linear with additive Gaussian noise applying the UKF in terms of the state update equations is straightforward. The observation model discussed above, however, involves Poisson distributed noise whose parameters depend upon the state and experimental settings. Thus, to apply the UKF, the model must be transformed into one where the measurement noise is Gaussian instead of Poisson. Two possible approaches are considered: One is a choice of a variance stabilizing transformation, such as the Anscombe or Freeman-Tukey transform, that yields a measurement model with additive Gaussian noise with unity variance (both are used here); the other is a straightforward replacement of the Poisson distribution by a Gaussian with a mean and variance equal to the rate of the original distribution.

The remainder of this paper is organized as follows. In Sec.~\ref{sec:problem}, we describe the problem formulation, including the motion and observation models in SPT application. Also, we describe the SPT application and introduce the motion and observation models used. This is followed in Sec.~\ref{sec:algorithm} by a description of the general U-EM technique. In Sec.~\ref{sec:sim} we use simulations to demonstrate the efficacy of our approach and to investigate the effect of the choice of transformation of the observation model under different experimental settings. Brief concluding remarks are provided in Sec.~\ref{sec:conc}.

\section{PROBLEM FORMULATION \label{sec:problem}}

\begin{figure}[htbp!]
\vspace{-0.2cm}
    \centering
    \includegraphics[width=3.3in]{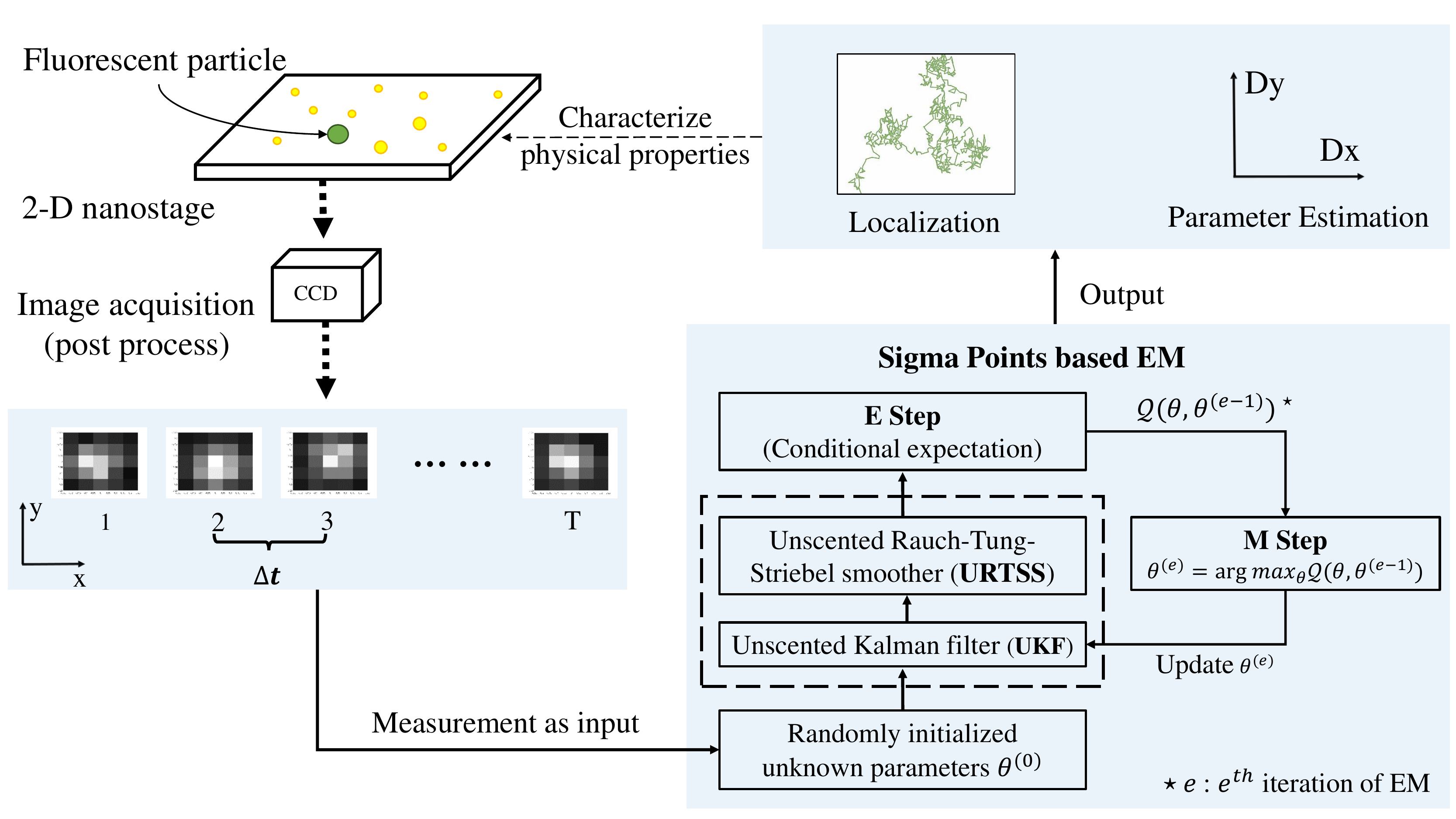}
    \vspace{-0.4cm}
    \caption{Generic framework of SPT study by Sigma Points based EM}
    \label{fig:generic_framework}
\end{figure}

The outline of our scheme is shown in Fig.~\ref{fig:generic_framework}. The left side of the figure represents the experimental techniques for acquiring data in which a particle of interest is labeled with a fluorescent tag (such as a fluorescent protein or quantum dot) and imaged through an optical microscope using a CCD camera. The image frames are then segmented to isolate individual particles. These segmented frames are then the input to the U-EM algorithm. In the remainder of this section we describe the motion and observation models used.

\vspace{-0.4cm}
\subsection{Motion Model}
For concreteness and simplicity of presentation, we focus on anisotropic diffusion in 2-D, though the extension to 3-D or to other common motion models (including directed motion, where the labeled particle is carried by the machinery of the cell, Ornstein-Uhlenbeck motion, which captures tethered motion, or confined diffusion) is straightforward. The model of anisotropic diffusion is
\begin{equation}
    X_{t+1}=X_t+\mathcal{N}(0,Q),\label{motion}
\end{equation}
where $X_t \in \real^2$ represents the location of the particle in the lateral plane at time $t$ and $Q$ is a covariance matrix given by
\begin{equation}
    Q=\left[\begin{array}{cc}
         2 D_x \Delta t & 0\\
         0 & 2 D_y  \Delta t   
         \end{array} \right].
\end{equation}
Here $D_x$ and $D_y$ are independent diffusion coefficients and $\Delta t$ is the time between frames of the image sequence.

\subsection{Observation Model} 
Because the single particle is smaller than the diffraction limit of light, the image on the camera is described by the point spread function (PSF) of the instrument. In 2-D (and in the focal plane of the objective lens), the PSF is well approximated by 
\begin{equation}
    PSF(x,y)={\exp}\left(-\frac{x^2}{2\sigma_{x}^{2}}-\frac{y^2}{2\sigma_{y}^{2}}\right),
\end{equation}
where $\sigma_{x}$ and  $\sigma_{y}$ are given by  
\begin{equation}
    \sigma_{x}=\sigma_y=\frac{\sqrt{2}\lambda}{2\pi \mathrm{NA}}.
\end{equation}
Here $\lambda$ is the wavelength of the emitted light and NA is the numerical aperture of the objective lens being used \cite{zhang2007gaussian}. This PSF is then imaged by the CCD camera.

Assuming segmentation has already been done (which is a standard pre-processing step), the image acquired by the camera is composed of $P$ pixels arranged into a $\sqrt{P}\times\sqrt{P}$ square array. The pixel size is $\Delta x$ by $\Delta y$ with the actual dimensions determined both by the physical size of the CCD elements on the camera and the magnification of the optical system. At time step $t$, the expected photon intensity measured for the $p^{th}$ pixel is then
\begin{equation*}
    \lambda_{p,t}=\int_{x_{p,t}^{min}}^{x_{p,t}^{max}}\int_{y_{p,t}^{min}}^{y_{p,t}^{max}} \frac{G}{\Delta x \Delta y} PSF(x_t-\xi,y_t-\xi')d\xi d\xi'
\end{equation*}
where $G$ denotes the peak intensity of the fluorescence and the integration bounds are over the given pixel.

In addition to the signal, there is always a background intensity rate arising from out-of-focus fluorescence and autofluorescence in the sample. This is typically modeled as a uniform rate $N_{bgd}$ \cite{ashley2015method}. Combining these signals and accounting for  the shot noise nature of the photon generation process, the measured intensity in the $p^{th}$ pixel at time $t$ is 
\begin{equation}
    I_{p,t}\sim \mathrm{Poiss}(\lambda_{p,t}+N_{bgd})\label{Ipt}.
\end{equation}
where Poiss($\cdot$) represents a Poisson distribution.
\subsection{Measurement Model Transformation}
The UKF is developed with an assumption of Gaussian-distributed noise \cite{sarkka2013bayesian}. We therefore need to transform the Poisson distributed model in \eqref{Ipt} into an appropriate form. We consider three possibilities.

\textit{Direct Gaussian Approximation:}
For a sufficiently high rate, a Poisson distribution of rate $\lambda$ is well approximated by a Gaussian of mean and covariance equal to that rate \cite{gnedenko2017theory}. One approach, then, is to replace \eqref{Ipt} with
\begin{align}
    \tilde{I}_{p,t} & =(\lambda_{p,t}+N_{bgd})+ v_k, \,
    v_k \sim \mathcal{N}\left(0,\lambda_{p,t}+N_{bgd}\right).
    \label{eq:gauss_approx}
\end{align}

This approach requires no modification to the measured data. However, the noise term $v_k$ itself depends upon the state variable since the rate $\lambda_{p,t}$ is a function of $X_t$.

\textit{Anscombe Transformation:} 
The Anscombe transformation is a variance-stabilizing transformation that (approximately) converts a Poisson-distributed random variable into a unit variance Gaussian one \cite{anscombe1948transformation}. Under this approach, the measurements are first transformed by
\begin{align}
    \tilde{I}_{p,t} = 2\sqrt{I_{p,t} + \frac{3}{8}}.
\end{align}
The measurement model \eqref{Ipt} is then replaced by
\begin{align}
    \tilde{I}_{p,t} \simeq 2\sqrt{\lambda_{p,t} + \frac{3}{8}} - \frac{1}{4\sqrt{\lambda_{p,t}}} + v_k, \, v_k \sim \mathcal{N}(0,1).
    \label{eq:anscombe}
\end{align}

\textit{Freeman Tukey Transformation:}
An alternative variance stabilizing transform is the Freeman and Tukey \cite{freeman1950}. Under this approach, the measurements are first transformed by
\begin{align}
    \tilde{I}_{p,t} = \sqrt{I_{p,t} + 1} + \sqrt{I_{p,t}}.
\end{align}
and the measurement model is replaced by
\begin{align}
    \tilde{I}_{p,t} \simeq \sqrt{\lambda_{p,t} + 1} + \sqrt{\lambda_{p,t}} + v_k, \, v_k \sim \mathcal{N}(0,1). \label{eq:freeman}
\end{align}


\section{Unscented Expectation Maximization \label{sec:algorithm}}

In this section we describe the U-EM approach which consists of the expectation maximization algorithm for finding an (approximate) maximum likelihood estimate of the parameters together with the UKF and URTSS for estimating the smoothed distribution of the latent variable (the trajectory of the particle in the SPT application).

\subsection{Parameter Estimation via Expectation Maximization}
Consider the problem of identifying an unknown parameter $\theta \in \mathbb{R}^{n_\theta}$ for the nonlinear state space model
\begin{subequations}
\begin{align}
    X_{t+1} &=f_t(X_{t}, w_{t}, \theta) \label{eq:generic_motion}\\
    Y_t &=h_t(X_t, v_{t}, \theta)
\end{align}\label{eq:generalized_model}
\end{subequations}
where the $X_t \in {\mathbb{R}}^{n}$, $Y_t \in {\mathbb{R}}^{{p}}$, and $w_t$ and $v_t$ are independent, identitically distributed white noise processes (not necessarily Gaussian).

The primary goal is to determine a maximum likelihood (ML) estimate of $\theta$ from the data $Y_{1:T} \triangleq \{Y_1,\dots,Y_T\}$ since that estimator is known to be asymptotically consistent and efficient. That is, we would like
\begin{equation}
    \hat{\theta}=\arg \max_{\theta} \log p_{\theta}(Y_{1:T})
    \label{eq:ML}
\end{equation}
where we have expressed the estimator using the log likelihood. However, it is often the case that $p_{\theta}(Y_{1:T})$ is unknown or intractable, and thus \eqref{eq:ML} cannot be solved directly.

The EM algorithm overcomes this challenge by taking advantage of the latent variables $X_{1:T}$ and seeks to optimize the complete log likelihood $L_{\theta}(X_{0:T},Y_{1:T})$, given by
\begin{align}
    &L_{\theta}(X_{0:T}, Y_{1:T})=\log p_{\theta}(Y_{1:T}|X_{0:T})+\log  p_{\theta}(X_{0:T})\nonumber\\
    &= \log p_{\theta}(X_0)+\sum_{t=1}^{T}\log p_\theta(X_{t}|X_{t-1})+\sum_{t=1}^{T} \log p_{\theta}(Y_{t}|X_t)\label{eq:log_joint}.
\end{align}

Unfortunately the latent state is not available, only the measurements $Y_{1:T}$. EM handles this by forming an approximation $Q(\theta,\theta^{(e)})$ of $L_{\theta}$ to achieve the minimum variance estimate of the likelihood given the observed data and an assumption $\theta^{(e)}$ of the true parameter value. This is of course given by the conditional mean
\begin{equation}
    \mathcal{Q}(\theta,\ \theta^{(e)})=\mathbb{E}_{\theta^{(e)}}\left[L_{\theta}(X_{0:T},Y_{1:T}) | Y_{1:T} \right].
    \label{eq:Qgen}
\end{equation}
Using \eqref{eq:log_joint} in \eqref{eq:Qgen} yields
\begin{align}
    \mathcal{Q}(\theta,\ \theta^{(e)})& =I_1(\theta, \theta^{(e)})+I_2(\theta, \theta^{(e)})+I_3(\theta, \theta^{(e)})\label{eq:Q}
\end{align}
where
\begin{subequations}
\begin{align}
    I_1(\theta, \theta^{(e)}) &=\mathbb{E}\left[\log p(X_0|\theta)|Y_{1:T},\theta^{(e)}\right],\\
    I_2(\theta, \theta^{(e)}) &=\sum_{t=1}^{T}\mathbb{E}\left[\log p(X_t|X_{t-1})|Y_{1:T}, \theta^{(e)}\right],\\
    I_3(\theta, \theta^{(e)})&=\sum_{t=1}^{T}\mathbb{E}\left[\log p(Y_t|X_t)|Y_{1:T}, \theta^{(e)}\right].
\end{align}\label{eq:Is}
\end{subequations}

The calculation of $\mathcal{Q}(\theta,\theta^{(e)})$ is called the \textit{Expectation (E) step} at the $e^{th}$ iteration. It has been shown \cite{dempster1977maximum} that any choice of $\theta^{(e+1)}$ such that $\mathcal{Q}(\theta^{(e+1)},\theta^{(e)})> \mathcal{Q}(\theta^{(e)},\theta^{(e)})$ also increases the original likelihood, that is $p_{\theta^{(e+1)}}(Y_{1:T}) > p_{\theta^{(e)}}(Y_{1:T})$. Thus, the expectation step is followed by a \textit{Maximization (M) step} to produce the next estimate of the parameter,
\begin{align}
    \theta^{(e+1)} = \arg\max_{\theta}\mathcal{Q}(\theta,\theta^{(e)}).
\end{align}

To implement the E step (that is, to calculate $\mathcal{Q}$) by carrying out the expectations in \eqref{eq:Is}, it is necessary to know the posterior densities $ p(X_t|Y_{1:T})$ and $ p(X_t,X_{t-1}|Y_{1:T})$. 
If the underlying model in \eqref{eq:generalized_model} is linear with Gaussian noise then these distributions are easily obtained \cite{gibson2005robust}. For nonlinear systems, however, there is no hope of any exact, analytical solution. Therefore, either some form of approximation or numerical approach must be used. Here we take an approximation approach and apply the UKF and URTSS.

\subsection{Unscented Kalman Filter}
The UKF was developed by Julier and Uhlman to capture (an approximation to) the mean and covariance of a nonlinear stochastic process without relying on the linearization approach of the EKF \cite{julier1997new}. More details can be found in many sources, such as \cite{sarkka2013bayesian}.

The UKF forms a Gaussian approximation of the filtering posterior distribution,
\begin{equation}
    p(X_t|Y_t)\simeq {\mathcal{N}}(\mathbf{m}_t,\mathbf{P}_t),\label{eq:posterior_filter}
\end{equation}
where mean and covariance are calculated as follows.

\noindent \textbf{Prediction step:} First calculate the $2n+1$ sigma points (where $n$ is the dimension of the state) according to
\begin{subequations}
    \label{eq:sigmapts}
     \begin{align}
    \mathcal{X}&_{t-1}^{(0)}={\mathbf{m}}_{t-1},\\
    \mathcal{X}&_{t-1}^{(i)}={\mathbf{m}}_{t-1}+ \sqrt{(n+\zeta)}\left[\sqrt{{\mathbf{P}}_{t-1}}\right]_{i},\\
    \mathcal{X}&_{t-1}^{(i+n)}={\mathbf{m}}_{t-1}- \sqrt{(n+\zeta)}\left[\sqrt{{\mathbf{P}}_{t-1}}\right]_{i}
    \end{align}
    \end{subequations}
for $i=1,\dots,n$. Here $\left[\cdot \right]_{i}$ denotes the $i^{th}$ column of the matrix, $\sqrt{A}$ is the matrix square root of $A$, and $\zeta$ is a scaling parameter defined by
    \begin{equation}
        \zeta=\alpha^2(n+\kappa)-n
    \end{equation}
where $\alpha$, $\beta$ and $\kappa$ allow the users to tune the algorithm performance \cite{arasaratnam2009cubature, wan2000unscented}. The sigma points are then propagated through the motion model
    \begin{equation}
        \hat{\mathcal{X}}_{t}^{(i)}=f({{\mathcal{X}}_{t-1}^{(i)}}), \ i=0,...,2n,
    \end{equation}
and then combined to produce the predicted mean and covariance at time $t$ given data up to time $t-1$ according to
    \begin{align}
        \mathbf{m}_{t}^{-}&=\sum_{i=0}^{2n}W_{i}^{(m)}\hat{\mathcal{X}}_{t}^{i}, \\
        \mathbf{P}_{t}^{-}&=\sum_{i=0}^{2n}W_i^{(c)}(\hat{\mathcal{X}}_{t}^{(i)}-\mathbf{m}_{t}^{-})(\hat{\mathcal{X}}_{t}^{(i)}-\mathbf{m}_{t}^{-})^T+Q_{t-1}.
    \end{align}
The weights are given by
    \begin{subequations}
    \label{eq:weights}
    \begin{align}
        W&_{0}^{(m)}=\frac{\zeta}{n+\zeta},\ W_{0}^{(c)}=\frac{\zeta}{n+\zeta}+(1-{\alpha}^2+\beta),\\
        W&_{i}^{(m)}=W_{i}^{(c)}=\frac{1}{2(n+\zeta)}, \ i=1,\dots,2n.
    \end{align}
    \end{subequations}

\noindent\textbf{Update and filter:} A new set of sigma points $\mathcal{X}_t^-$ are formed from the predicted mean and covariance according to \eqref{eq:sigmapts} using $\mathbf{m}_t^-$ and $\mathbf{P}_t^-$ in lieu of $\mathbf{m}_{t-1}$ and $\mathbf{P}_{t-1}$.
These sigma points are then propagated through the measurement
    \begin{equation}
        \hat{\mathcal{Y}}_{t}^{(i)}=h(\mathcal{X}_{t}^{-(i)}), \ i=0,...,2n,
    \end{equation}
and combined to form
    \begin{align}
        \mu_t &=\sum_{i=0}^{2n} W_{i}^{(m)}\hat{\mathcal{Y}}_{t}^{(i)},\\
        S_t &=\sum_{i=0}^{2n}, W_{i}^{(c)}(\hat{\mathcal{Y}}_{t}^{(i)}-\mu_k)(\hat{\mathcal{Y}}_{t}^{(i)}-\mu_k)^T+\mathbf{R}_t, \\
        C_t &=\sum_{i=0}^{2n} W_{i}^{(c)}(\mathcal{X}_{t}^{-(i)}-\mathbf{m}^-)(\hat{\mathcal{Y}}_{t}^{(i)}-\mu_t)^T.
    \end{align}
where $\mathbf{R}_t$ is a covariance matrix in measurement model. Finally, these are used to produce the filtered estimates of the mean and covariance of the process at time $t$ using the data up to time $t$ through
    \begin{align}
        K_t &=C_t S_{t}^{-1},\\
        \mathbf{m}_t &= \mathbf{m}_t^{-}+K_t\left[Y_t-\mu_t \right],\\
        \mathbf{P}_t &= \mathbf{P}_t^{-}-K_t S_t K_t{}^T.
    \end{align}
    
\subsection{Unscented Rauch-Tung-Striebel Smoother}
To obtain (an approximation to) the distribution $p(X_t|Y_{1:T})$, we apply the URTSS \cite{4484208}. The URTSS  begins with the final results of the UKF, $\mathbf{m}_{T}^{s}=\mathbf{m}_T$ and $\mathbf{P}_{T}^{s}=\mathbf{P}_T$, and then runs a backward recursion from $t=T-1,...,0.$ as follows.

\noindent\textbf{Prediction and update:} First form the sigma points $\mathcal{X}_t$ from \eqref{eq:sigmapts} using $\mathbf{m}_t$ and $\mathbf{P}_t$.
These are then propagated through the motion model
    \begin{equation}
        \hat{\mathcal{X}}_{t+1}^{(i)}=f({{\mathcal{X}}_{t}^{(i)}}), \ i=0,1,...,2n
    \end{equation}
and combined to form
    \begin{align}
        \mathbf{m}_{t+1}^{-}&=\sum_{i=0}^{2n}W_{i}^{(m)}\hat{\mathcal{X}}_{t+1}^{(i)},\\
        \mathbf{P}_{t+1}^{-}&=\sum_{i=0}^{2n}W_i^{(c)}(\hat{\mathcal{X}}_{t+1}^{(i)}-\mathbf{m}_{t+1}^{-})(\hat{\mathcal{X}}_{t+1}^{(i)}-\mathbf{m}_{t+1}^{-})^T+Q_{t},\\
        D_{t+1}&=\sum_{i=0}^{2n}W_{i}^{(c)}(\mathcal{X}_t^{(i)}-\mathbf{m}_t)(\hat{\mathcal{X}}_{t+1}^{(i)}-\mathbf{m}_{k+1}^{-})^T,
    \end{align}
where the weights are given in \eqref{eq:weights}.

\noindent\textbf{Calculate the smoothed estimate:} The mean and covariance defining the smoothed Gaussian density at time $t$ are calculated from
    \begin{align}
        \mathcal{G}_t &=D_{k+1}\left[P_{t+1|T}^{-} \right]^{-1},\\
        \mathbf{m}_{t|T}^{s} &=\mathbf{m}_{t}+\mathcal{G}_t(m_{t+1|T}^{s}-\mathbf{m}_{t+1}^{-}),\\
        \mathbf{P}_{t|T}^{s} &= \mathbf{P}_{t}-\mathcal{G}_t(\mathbf{P}_{t+1|T}^{s}-\mathbf{P}_{t+1}^{-})\mathcal{G}_t^T.
    \end{align}

From the UKF and URTSS, we form the approximated posterior densities needed for the EM algorithm
\begin{align}
    p(X_t|Y_{1:T}) &\sim \mathcal{N} (\mathbf{m}^s_{t|T},\mathbf{P}^s_{t|T}), \label{posteriorPorb}\\
    p(X_t,X_{t-1}|Y_{1:T}) &\sim \nonumber \\
    &\hspace{-1.25cm}\mathcal{N}\left( \begin{bmatrix} \mathbf{m}^s_{t|T} \\ \mathbf{m}^s_{t-1|T} \end{bmatrix},\begin{bmatrix} \mathbf{P}^s_{t|T} & \mathbf{P}^s_{t|T}\mathcal{G}^T_{t-1} \\
    \mathcal{G}_{t-1}\mathbf{P}^s_{t|T} &  \mathbf{P}^s_{t-1|T} \end{bmatrix}\right).
    \label{jointPostProb}
\end{align}

\subsection{Applying U-EM to the SPT Setting}

Applying U-EM is primarily a matter of identifying the specific model for \eqref{eq:generalized_model} and the parameters to be identified. As we are focusing on anisotropic diffusion, the motion model is given by \eqref{motion} which depends on unknown diffusion coefficients. The observation model depends on the choice of transformation and is given either by \eqref{eq:gauss_approx}, \eqref{eq:anscombe}, or \eqref{eq:freeman}. The unknown parameters are diffusion coefficients $D_x$ and $D_y$. 
\section{DEMONSTRATION AND ANALYSIS \label{sec:sim}}

To demonstrate the performance of the U-EM algorithm in the SPT setting, we performed several simulations. 40 different ground truth trajectories were generated from the diffusion motion model \eqref{motion} and used to create simulated images according to the observation model in \eqref{Ipt}. The optical parameters and other fixed constants used in these simulations are shown in Table \ref{table_para}; these were chosen to mimic experimental settings found in many SPT experiments. 
\vspace{-0.4cm}
 \begin{table}[htbp!]
  \caption{Parameter settings}
  \vspace{-0.5cm}
  \begin{center}
    \begin{tabular}{|c||c||c|}
    \hline
    \rowcolor{gray}
    \textcolor{white}{Symbol} & \textcolor{white}{Parameter} & \textcolor{white}{Values} \\
    \hline
    \hline
    $\Delta t $ & Image period (discrete time step) & $100\ \text{ms}$ \\
   \hline
    $T$& Number of images per dataset & $100$\\
    \hline
    $P$& Number of pixels per squared image & $25$\\
    \hline
    $D_x$  & Diffusion coefficient in x direction & $0.005\ \rm{\mu m^2/s}$\\
    \hline
    $D_y$  & Diffusion coefficient in y direction & $0.01\ \rm{\mu m^2/s}$\\
    \hline
    $\Delta x$ &Length of unit pixel  & $100 \ \text{nm}$ \\
    \hline
    $\Delta y$ & Width of unit pixel  & $100 \ \text{nm}$ \\
    \hline
    $\lambda$ & Emission wavelength & $540\ \text{nm}$ \\
    \hline
   $ NA$    & Numerical aperture & $1.2$ \\
    \hline
    \end{tabular}%
  \label{table_para}%
  \end{center}
 \vspace{-1cm}
\end{table}%

\subsection{Demonstration}

\begin{figure}[htbp!]
        \centering\includegraphics[width=0.35\linewidth]{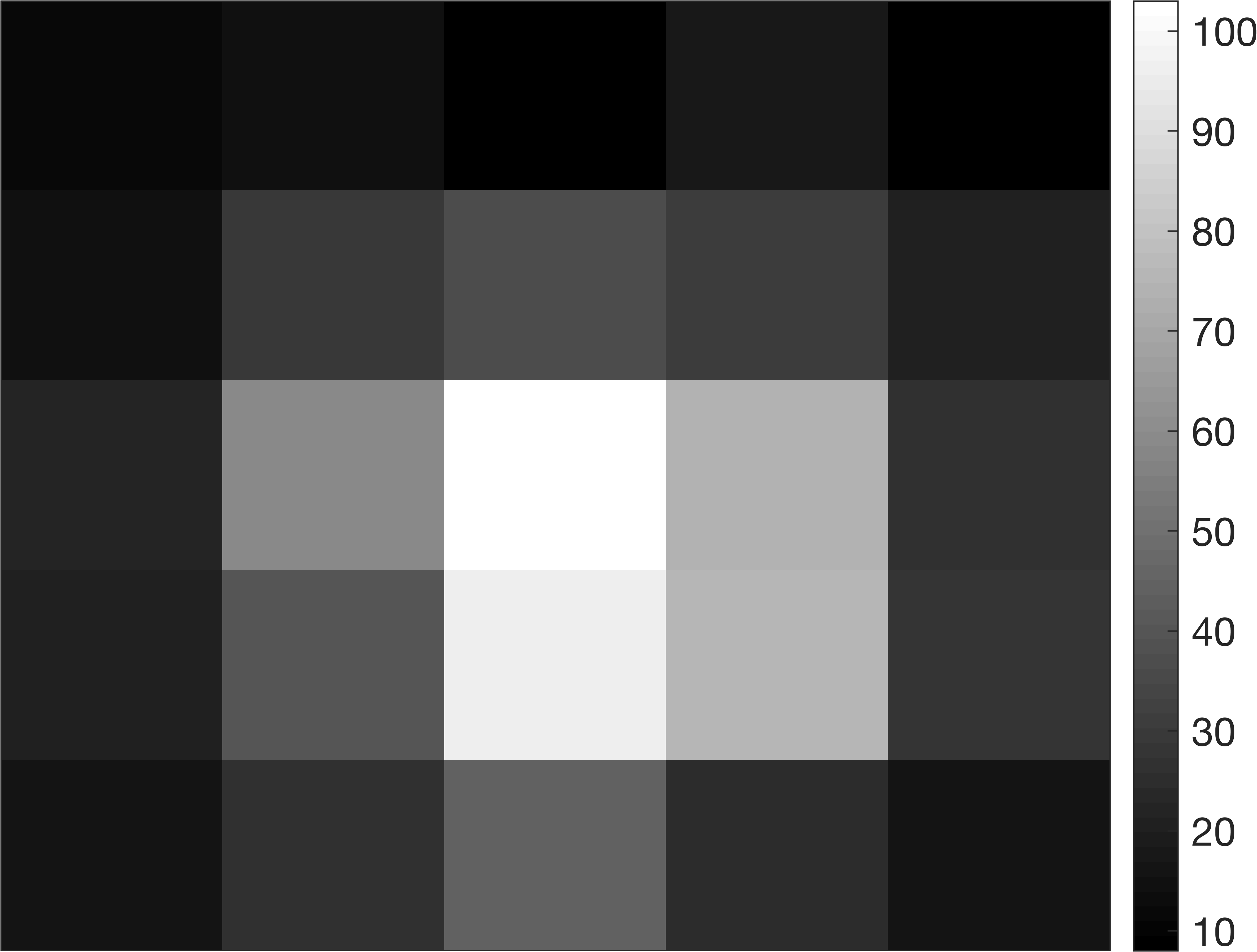}
        \includegraphics[width=0.34\linewidth]{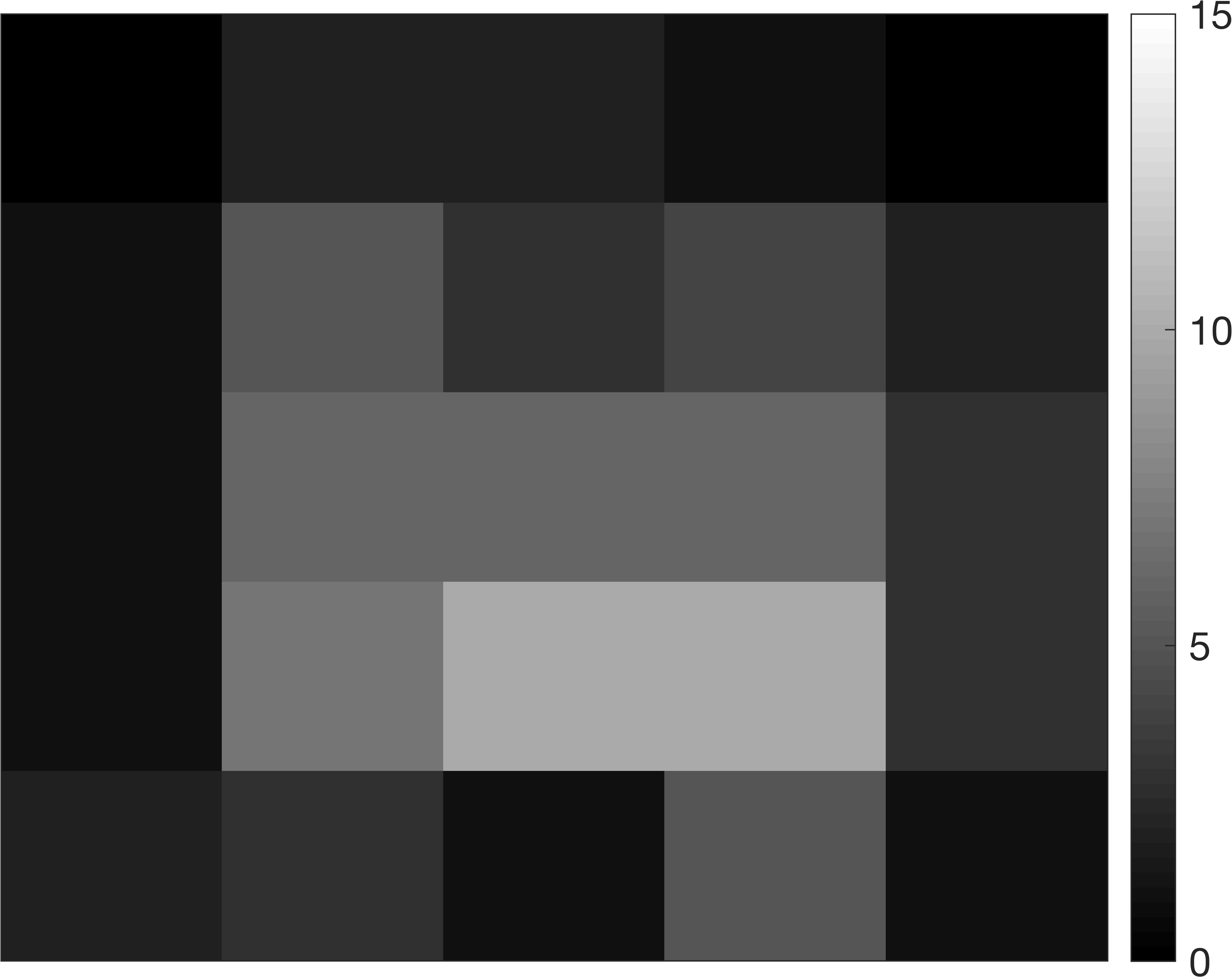}
    \caption{Typical data images with (left) $N_{bgd}=10$ and $G=100$ and (right) $N_{bgd}=1$ and $G=10$. There are a total of 867 photon counts captured among the 25 pixels in the left image and 85 counts in the right image. Notice the different scaling in the two images.}
    \label{fig:N10G100}
    \vspace{-0.4cm}
\end{figure}

\begin{figure}[htbp!]
    \centering
    \includegraphics[width=1\linewidth]{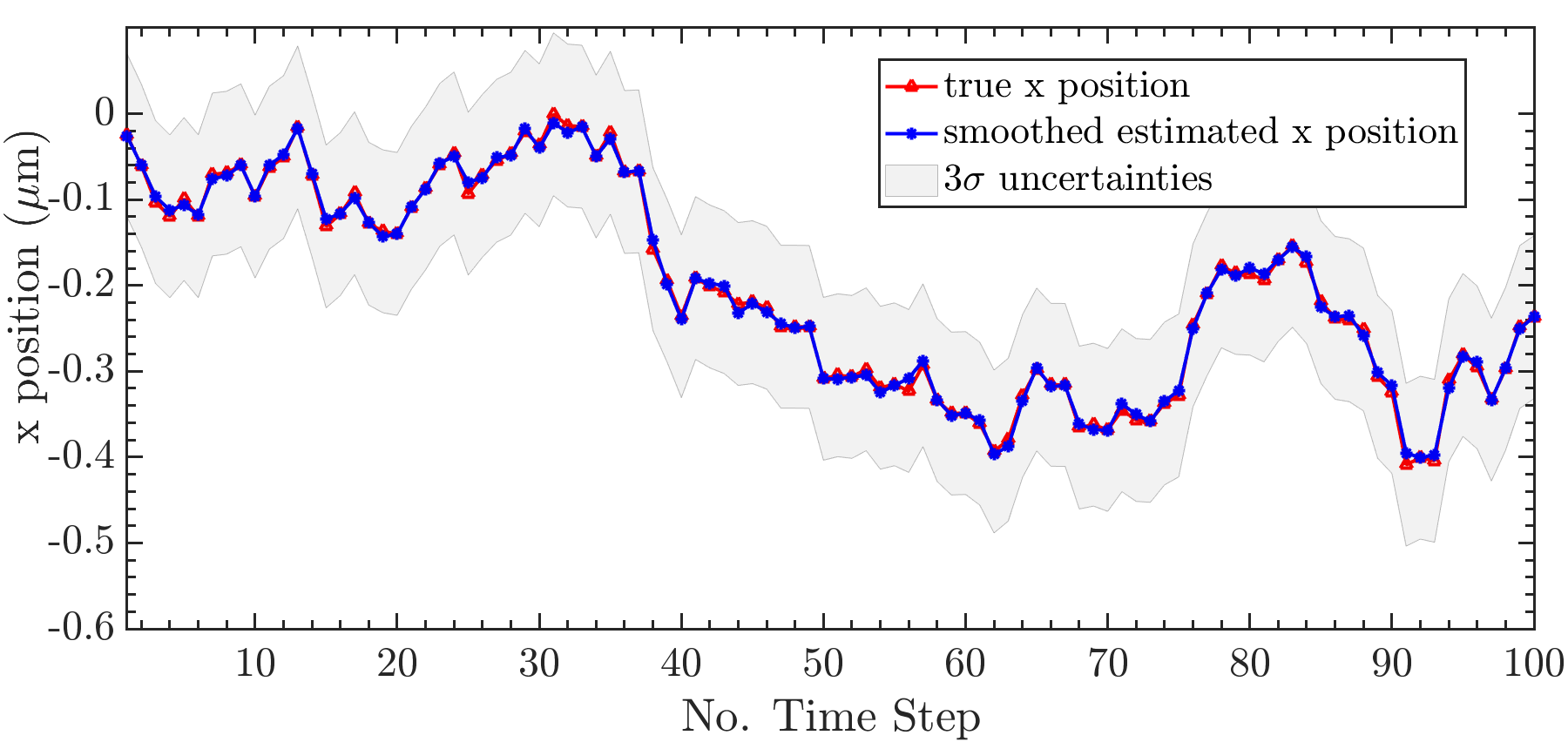}
    \vspace{-0.4cm}
\caption{Position estimation through Anscombe transform in $x$ direction}
    \label{fig:estimation_x}
    \vspace{-0.4cm}
\end{figure}

To demonstrate, we fixed the background rate $N_{bgd} = 10$ and the peak signal intensity $G=100$, representing a strong but not atypical signal in actual SPT experiments \cite{ashley2015method,chenouard2014objective}. A typical image is shown in Fig.~\ref{fig:N10G100}. The algorithm was applied across 40 sample trajectories. A typical trajectory estimation result, calculated using the Anscombe transform to the measurement model, is shown in Fig.~\ref{fig:estimation_x}. One interesting feature of the U-EM approach is that the trajectory estimation yields a (Gaussian) distribution at each time step rather than a single point estimate. In Fig.~\ref{fig:estimation_x}, the results show the mean tracks the true path very closely with a tight distribution. 
\vspace{-0.3cm}
\vspace{-0.3cm}

\begin{figure}[htbp!]
    \centering
    \includegraphics[width=0.9\linewidth]{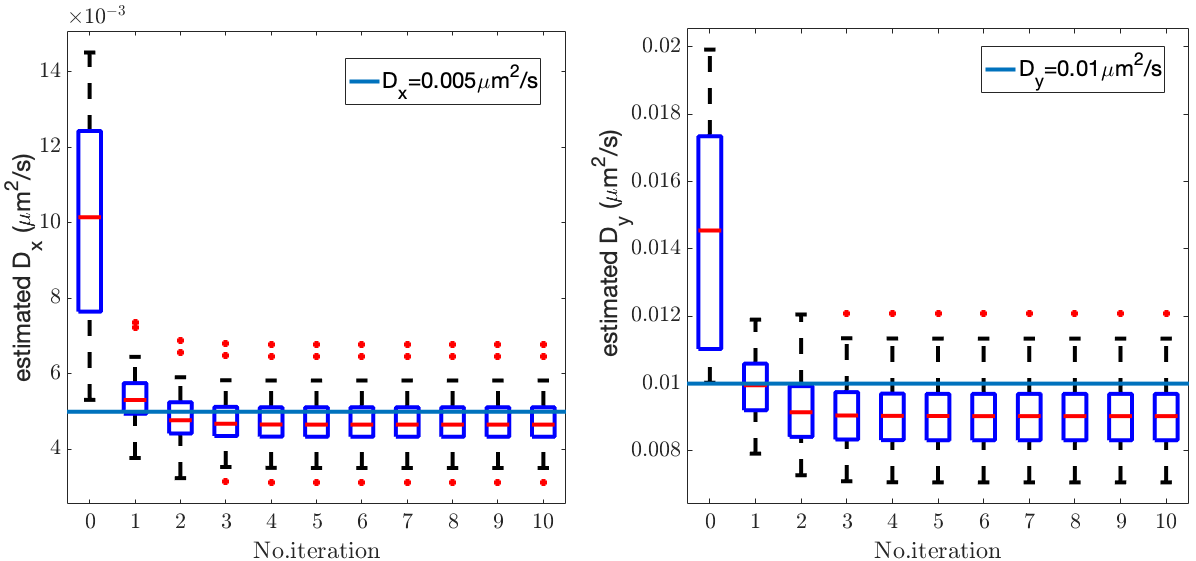}
    \vspace{-0.4cm}
\caption{Estimation of $D_x$ and $D_y$ using the Gaussian approximation to the measurement model. As with all box plots, the (red) line in the box denotes the median, the edges of the box show the first and third quartiles, the vertical dashed lines indicate bounds of 1.5 times the interquartile range, and the red dots indicate outliers.}
    \label{fig:est_Dxy_gauss}
    \vspace{-0.3cm}
\end{figure}

The evolution of the diffusion coefficient estimate as a function of EM iteration is shown in Fig.~\ref{fig:est_Dxy_gauss}. (These estimates were done using the Gaussian approximation to the measurement model.) These results show rapid convergence to a value quite close to the true diffusion coefficients. 
\vspace{-0.3cm}

\begin{figure}[htpb!]
    \centering
    \includegraphics[width=0.9\linewidth]{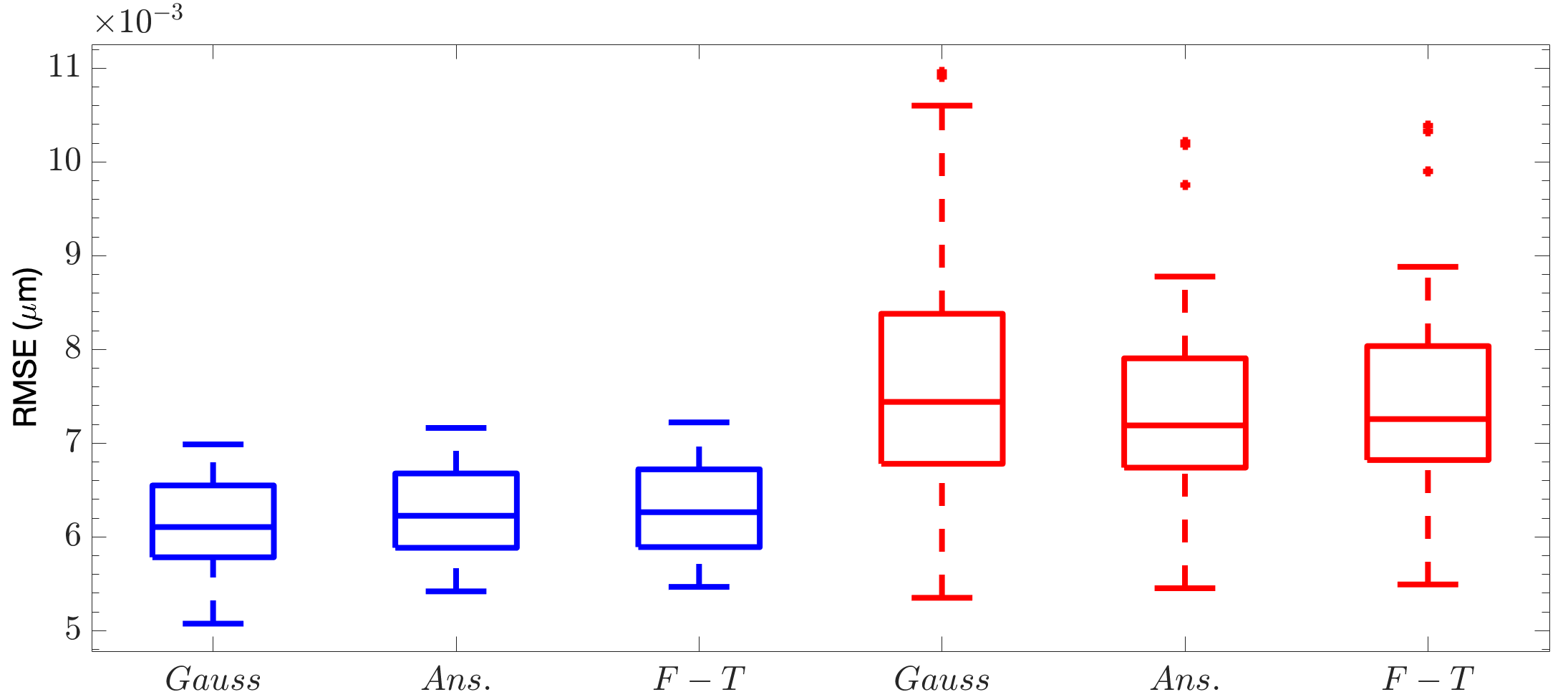}
    \vspace{-0.2cm}
\caption{Box plots of 2-D position estimation error using the (Gauss) Gaussian approximation,  (Ans.) Anscombe  transform, and (F-T) Freeman-Tukey transform. Blue and red box correspond to RMSE in x and y position respectively. }
    \label{fig:boxplot_rmse_xy}
\end{figure}

To explore the difference among these transformations, the simulations were repeated with each of the three choices. The estimated position at each time was taken as the mean value of the smoothed distribution. The resulting root mean square errors (RMSE) are shown in Fig.~\ref{fig:boxplot_rmse_xy}. As can be seen, all approaches perform well with an estimation error of approximately 6.25\ nm in $x$ position and 7.30\ nm in $y$ position. Both the similarity and the actual error level is as expected given that the signal level is high.
\par Performance of parameter estimation over the 40 runs and with the three different transformation choices is shown in Table~\ref{tab:est_Dxy}. 
\begin{table}[htbp!]
  \centering
  \caption{Parameter estimation of $D_x$ and $D_y$ on 40 datasets}
    \vspace{-0.3cm}
    \begin{tabular}{|l||c||c|}
    \hline
    \rowcolor{gray}
    \textcolor{white}{Method} & \textcolor{white}{$D_x$ ($\mu$m$^2$/s)} & \textcolor{white}{$D_y$ ($\mu$m$^2$/s)} \\
    \hline
    \hline
    Gaussian & 0.0047 $\pm$ 7.3e-4 & 0.009 $\pm$ 0.0011 \\
    \hline
    Anscombe & 0.0046 $\pm$ 7.3e-4 & 0.009 $\pm$ 0.0011 \\
    \hline
    Freeman Tukey & 0.0046 $\pm$ 7.3e-4 & 0.009 $\pm$ 0.0011 \\
    \hline
    \end{tabular}%
  \label{tab:est_Dxy}%
  \vspace{-0.5cm}
\end{table}%

\subsection{Performance across Different Signal Levels}
The primary differences among the different observation model transformations become meaningful only when the rate of the Poisson distribution is low (as determined by the combination of signal level and background). We performed two sets of simulations at different noise levels \cite{chenouard2014objective}. In the first set, the noise $N_{bgd}$ was fixed at one and the signal $G$ increased from one to 10. In the second, Signal to Noise Ratio (SNR) was fixed to 10 and $N_{bgd}$ increased from 1 to 10. Other imaging and model parameters were kept the same.

\begin{figure}[htbp!]
    \centering
    \includegraphics[width=0.9\linewidth]{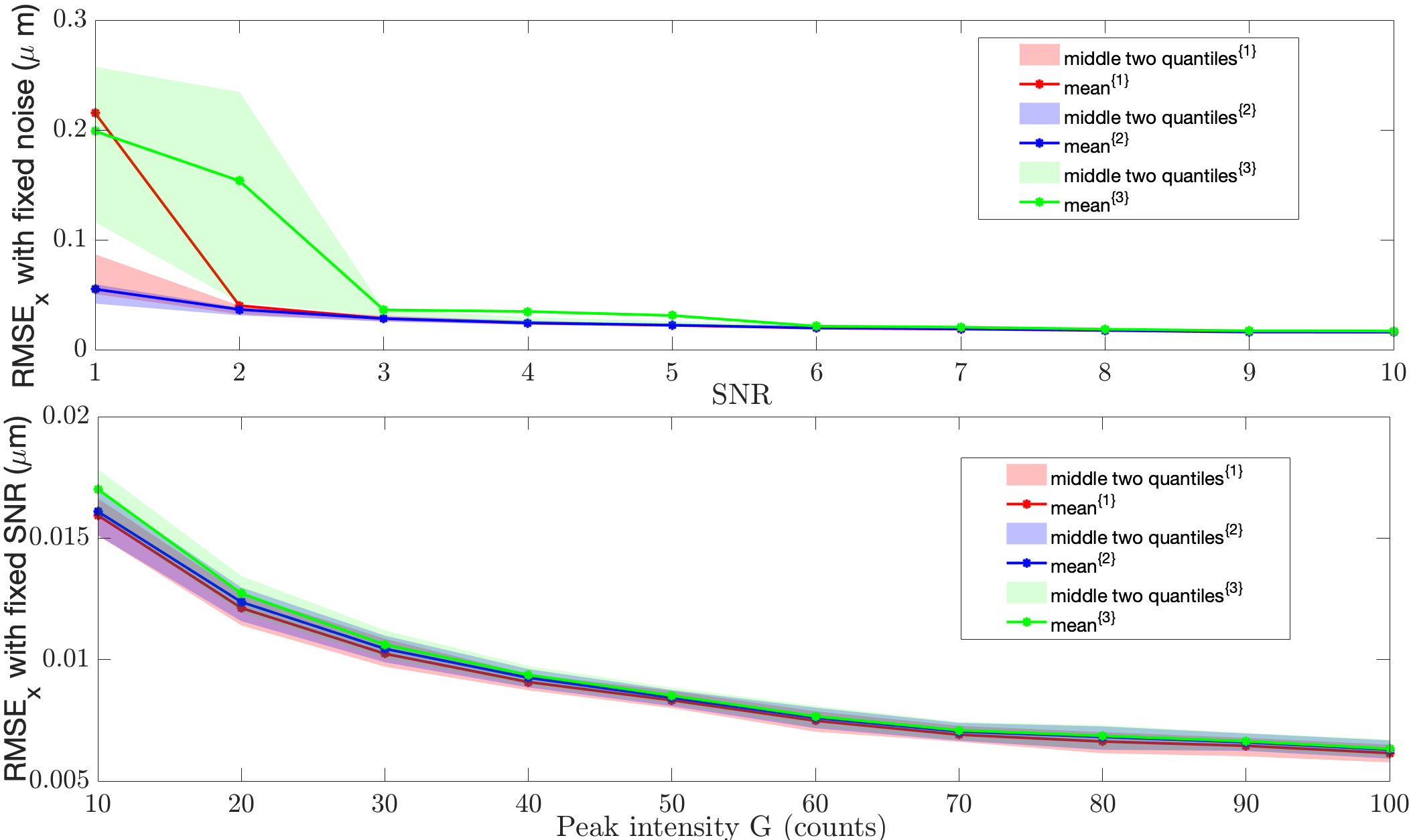}
    \vspace{-0.4cm}
\caption{RMSE of x position estimation with different $\{N_{bgd}, \ G\}$. The superscript \{1\}, \{2\} and \{3\} indicates results based on the Gaussian approximation,  Anscombe  transform, and  Freeman-Tukey  transform,  respectively. (top) Results holding $N_{bgd}$=1 fixed and varying $G$, showing the behavior at very low signal levels. (bottom) Results holding the ratio $SNR=10$ fixed while varying $G$. Note that for space reasons, only results for $D_x$ are shown; estimation of $D_y$ is similar.}
    \label{fig:rmse_logSpace}
    \vspace{-0.4cm}
\end{figure}

The localization results are shown in Fig.~\ref{fig:rmse_logSpace} with the top graph corresponding to $N_{bgd}=1$ (and thus extremely low signal levels) and the bottom to simulations for a fixed SNR. In both plots, red corresponds to Gaussian approximation, blue to Anscombe transform and green to Freeman-Tukey transform. It is clear that differences only appear at the low signal levels. Note that in the first plot with a peak intensity of $G=6$, the rate in the pixel at the center of the PSF is still only 7 counts. At the lowest signal levels, the Anscombe transform outperforms the other two. While the Gaussian approach is close, the difference between the mean and the center quartiles indicates that it has several large outliers. To put these estimation errors in context, note that for the given imaging parameters, the diffraction limit of light is approximately 270 nm.

\begin{figure}[htbp!]
    \centering
    \includegraphics[width=0.9\linewidth] {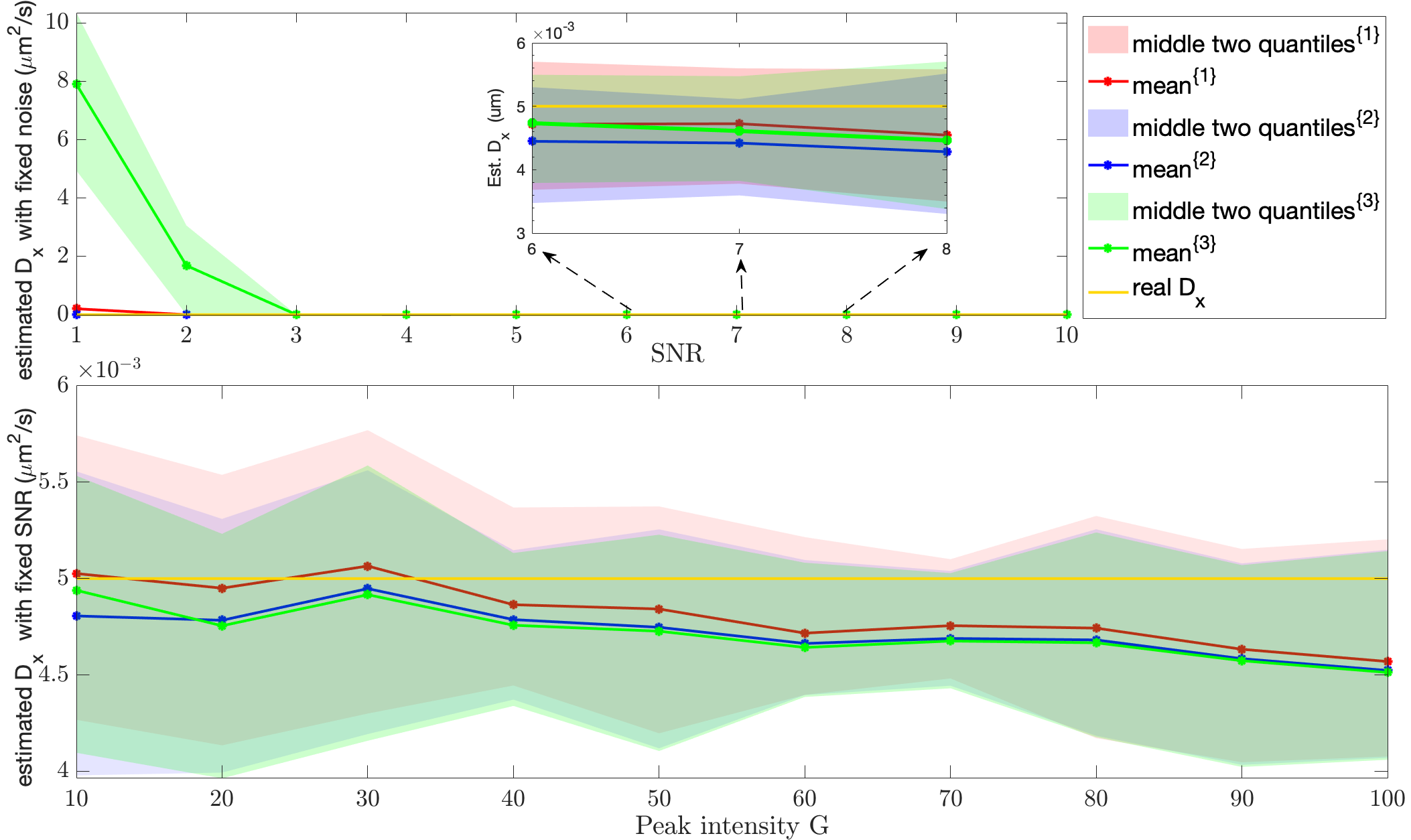}
    \vspace{-0.3cm}
\caption{Results of estimation of $D_x$ with a true value of $D_x = 0.005$ $\mu$m/s$^2$. (top) Results holding $N_{bgd}=1$ fixed while varying $SNR$. (bottom) Results holding the ratio $SNR$ fixed while varying $G$.}
\vspace{-0.3cm}
    \label{fig:Dxy_EM_6_transform}
\end{figure}

The corresponding results for the estimation of the diffusion coefficients are shown in Fig.~\ref{fig:Dxy_EM_6_transform}. As before, red corresponds to Gaussian approximation, blue to Anscombe transform, and green to Freeman-Tukey transform. The true value is $D_x = 0.005$ $\mu$m/s$^2$. These results parallel the trajectory estimation, with all transformations of the observation equation being essentially equivalent at high signal levels and Freeman-Tukey failing at the lowest signals. 
\addtolength{\textheight}{-2cm}

\par As noted in Sec.~\ref{sec:intro}, compared with SMC-EM, the method is faster with all of the transformation methods (on the order of a few minutes using unoptimized code in Matlab on a standard laptop with 10 EM iterations), the Gaussian approximation runs the slowest of the three (approximately 10-15\% slower). This is easily explained from the equations \eqref{eq:gauss_approx} - \eqref{eq:freeman} where we see that under the Gaussian approximation, the variance $\lambda_{p,t}$ must be calculated at each time step while both Anscombe and Freeman-Tukey avoid this since they are variance stabilized to one. Since the Anscombe transform outperforms at low signal level and has lower computational load, it should be the preferred approach. 

\section{CONCLUSIONS \label{sec:conc}}

In this paper the U-EM algorithm is introduced to the application of localization and parameter estimation in SPT. We explored the use of three different transformation methods to bring the observation model describing the camera images in SPT into a form amenable to the UKF, namely using a direct Gaussian approximation of the Poisson-distributed random variable modeling the intensity measurements on the camera and transforming the measurements using an Anscombe or Freeman-Tukey transform to convert them into unity variance, Gaussian distributed random variables. At high signal levels, all three approaches produce similar results but that at very low signal levels, the Anscombe outperforms the others (though with the Gaussian approximation close behind). In future work we plan to incorporate other, biologically relevant motion models, as well as introduce additional complexities into the observation model to capture, for example, camera-specific noise.

\section*{ACKNOWLEDGEMENT}
The authors gratefully acknowledge B. Godoy and N. A. Vickers  for insightful discussions. This work was supported in part by NIH through 1R01GM117039-01A1.
\bibliographystyle{IEEEtran}
\bibliography{IEEEabrv,bibliography}
\end{document}